\documentstyle [12pt,epsf] {article}
\def\TD{$T$}
\def\PD{$P$}
\def\CD{$C$}
\def\EWP{$P_{EW}$}
\def\AD{$A$}
\def\EX{$E$}
\def\PA{$PA$}
\def\etal{{\it et al.}}

\def\Journal#1#2#3#4{{#1} {\bf #2}, #3 (#4)}

\def\NIM{\em Nucl. Instrum. Methods}

\def\NPB{{\em Nucl. Phys.} B}
\def\PLB{{\em Phys. Lett.}  B}
\def\PRL{\em Phys. Rev. Lett.}
\def\PRD{{\em Phys. Rev.} D}
\def\ZPC{{\em Z. Phys.} C}

\def\be{\begin{equation}}
\def\ee{\end{equation}}
\def\bea{\begin{eqnarray}}
\def\eea{\end{eqnarray}}

\begin{document}
\begin{flushright}
CALT-68-2121\\
\today 
\end{flushright}

\vspace*{3.5cm}
\centerline{\large\bf B Decays to Two Charmless Pseudoscalar Mesons at CLEO
\footnote{
Presented at XXXIInd Rencontres de Moriond on ``QCD and High Energy Hadronic 
Interactions'', 1997}
}
\vspace*{6.0ex}
\centerline{\large Frank W\"urthwein}
\vspace*{1.5ex}
\centerline{\large\it Caltech}
\centerline{\large\it m/c 256-48}
\centerline{\large\it Pasadena, CA 91125, USA}
\centerline{\large\it email: fkw@cithe502.cithep.caltech.edu}
\vspace*{4.5ex}
\vfill
\begin{abstract}
Using $3.3$\ Million $B\bar B$\ pairs accumulated with the CLEO
detector we have measured 
${\cal B}(B^0\to K^+\pi^-) = 
(1.5^{+ 0.5}_{-0.4}\pm 0.1\pm 0.1)\times 10^{-5}$,   
${\cal B}(B^+\to K^0\pi^+) = 
(2.3^{+1.1}_{-1.0}\pm 0.2\pm 0.2)\times 10^{-5}$,   
and
${\cal B}(B^+\to \eta^{'}K^+) = 
(7.8^{+ 2.7}_{-2.2}\pm 1.0)\times 10^{-5}$.
These constitute the first observations of exclusive $B$\ decays to charmless 
hadronic final states.
Furthermore, a measurement of
${\cal B}(B^+\to h^+\pi^0) = 
(1.6^{+ 0.6\ +0.3}_{-0.5\ -0.2}\pm 0.1)\times 10^{-5}$, 
as well as upper limits on various
other $B$\ decays to two charmless pseudoscalar mesons are presented.
In particular, an upper limit of
${\cal B}(B^0\to \pi^+\pi^-) <
1.5\times 10^{-5}$ @ $90\%$C.L. is placed.
All of these results are still preliminary,
and averaging over charge conjugate 
modes is always implied.
\end{abstract}
\vfill
\ \ \ \ 
\newpage
\section{Introduction}
To lowest order,
hadronic $B$ decays can be described by external (\TD) and internal (\CD) 
W-emission,
gluonic (\PD) and electroweak (\EWP) penguins, as well as annihilation
(\AD), W-exchange (\EX), and penguin annihilation (\PA) diagrams.
Neglecting CKM-matrix elements one might naively expect 
\PD/\TD$\approx O(\alpha_s(m_b))\approx 0.2$, \CD/\TD$\approx 1/3$, and
\EWP/\PD
$\approx 10\% $~\cite{gronauewp}. 
\AD, \EX, \PA\ ought to be very small compared to \TD\ 
as they are suppressed by $f_B/m_B \approx 5\% $. Taking CKM matrix
elements into account Gronau \etal\cite{gronauewp}\ 
have suggested an approximate hierarchy in orders of $\lambda\approx 0.2$
as follows:
\begin{equation}
\begin{array}{ccc}
 & \Delta S=0 & \Delta S=1 \\
1 & T & P \\
\lambda & C,P & T, P_{EW} \\
\lambda^2 & E,A,P_{EW} & C,PA,P^C_{EW} \\
\lambda^3 & PA,P^C_{EW} & E,A
\end{array}
\end{equation} 

\begin{figure}
\centering
\leavevmode
\epsfysize=7cm
\epsfxsize=12cm
\epsfbox[44 233 467 538]{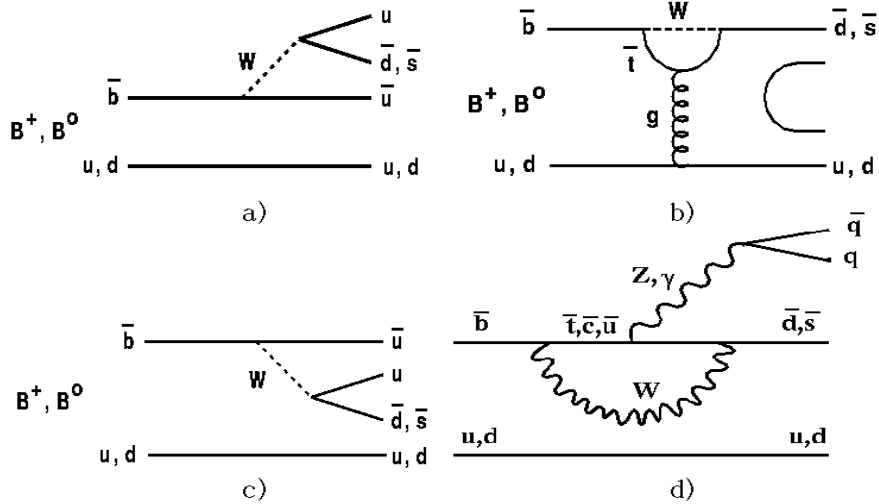}
\caption{The dominant decay processes are expected to be 
a) external W-emission, b)gluonic Penguin, c) internal W-emission,
d) external electroweak Penguin.}
\label{fig:diagrams}
\end{figure}

$P^C_{EW}$ denotes the internal electroweak penguin diagram which 
is color suppressed. Gronau \etal\ assume that \PD\ is dominated by
$t-quark$\ loop. \PD$=P_t$\ is then suppressed by $|V_{td}/V_{ts}|$\ in
$b\to d$\ as compared to $b\to s$\ penguin amplitudes. 
Fleischer and Mannel
~\cite{flpeng}\ suggested $P_c/P_t\approx 0.6-0.7$\ for 
$b\to s$\ penguin amplitudes. A similar $O(1)$\ ratio may be expected for
$b\to d$\ penguins.
Figures~\ref{fig:diagrams}a) to d) depict the four dominant diagrams.

The CLEO II experiment~\cite{cleo}
has accumulated $3.3\times 10^6\ B\bar B$\ pairs.
With typical efficiencies for two-body final states of $20-40\%$, and
backgrounds of only a fraction of an event per million $B\bar B$\ pairs
we are sensitive to branching fractions as low as a few times $10^{-6}$,
in some cases. Predictions for the dominant \TD\ and \PD\ amplitudes 
translate
into branching fractions at a level of one to few times $10^{-5}$.
We are therefore in a position to provide first experimental tests
of theoretical predictions for absolute~\cite{pengprediction}, 
as well as ratios of branching fractions~\cite{relprediction}.

Many authors have proposed to use charmless hadronic $B$\ decays to
probe the CKM sector of the standard model. 
For a recent review of this topic
see for example Ref.~\cite{fleischerreview}.
An experimental test of the hierarchy of decay amplitudes presented above
is crucial to assess the experimental and theoretical feasibility
of probing the standard model in this manner.

\section{Experimental Results}
The Cornell Electron Storage Ring (CESR) is
a symmetric 
$e^+e^-$\ collider operating at a center of mass energy near
the $\Upsilon$\ resonances. 
The hadronic cross section for
continuum production of u, d, s, or c quark anti-quark pairs
is about a factor of 3 higher than that for $e^+e^-\to\Upsilon(4S)\to
B\bar B$.
This continuum production is
the dominant background to the $B$\ decays of interest here.

The CLEO II detector~\cite{cleo} boasts excellent charged and neutral
particle detection. 
The momenta of charged particles is measured in a tracking system consisting 
of
a 6-layer straw tube chamber,
a 10-layer precision drift
chamber, and a 51-layer main drift chamber, all operating inside a 1.5 T
superconducting solenoid. The main drift chamber also provides
a measurement of dE/dx used for particle identification. Photons are detected
using 7800 CsI crystals, which are also inside the solenoid.
The return yoke is instrumented at various depths with proportional counters
to identify muons.
CLEO II has accumulated a grand total of $3.3$\ million $B\bar B$\ pairs.

Table~\ref{tab:results} lists the measured branching fractions and 
upper limits for
charmless hadronic $B$\ decays to
$\pi\pi ,\ K\pi ,\ KK$\ as well as final states containing
an $\eta$\ or $\eta^{'}$. 
This is an update to a previously published analysis~\cite{rarebprd}.
The main difference being a $30\%$\ increase in data, and loosening
of continuum background suppression cuts to increase the efficiency
by $\approx 12\% $.
Furthermore, we have extended the analysis to look for
modes containing $\eta$\ or $\eta^{'}$.

\begin{table}
\begin{center}
\caption{
Summary of  PRELIMINARY CLEO results for $B\to 
K\pi, \pi\pi, KK, h^+\eta, h^+\eta^{'}$. 
The results for $K^0\pi^0$\ and $\pi^0\pi^0$\ are 
CLEO(1995) Results (PRD 53, 1039 (1996)). An update for these
final states is still in progress.}
\label{tab:results}
\vskip 0.4cm
\begin{tabular}{llllll} 
\hline
Mode & Eff (\%) & Yield & Signif & BR ($10^{-5}$) & UL ($10^{-5}$)\\
\hline
 {$K^\pm\pi^\mp$}                          & 
 {44}                                      & 
 {$21.7_{-6.0}^{+6.8}$}                    & 
 {5.6$\sigma$}                             & 
 {$1.5^{+0.5+0.1}_{-0.4-0.1} \pm 0.1$}     & 
 {}\\
 {$K^\pm\pi^0$}                            & 
 {37}                                      & 
 {$8.7_{-4.2}^{+5.3}$}                     & 
 {2.7$\sigma$}                             & 
 {$0.7\pm 0.4$}                                        & 
 {1.6}\\
 {$K^0\pi^\pm$}                            & 
 {12}                                      & 
 {$9.2_{-3.8}^{+4.3}$}                     & 
 {3.2$\sigma$}                             & 
 {$2.3^{+1.1+0.2}_{-1.0-0.2} \pm 0.2$}     & 
 {4.4}\\                                    
 {$K^0\pi^0$} *                            & 
 {7}                                     & 
 {$2.3^{+2.2}_{-1.5}$}                     & 
 {}                                        & 
 {}                                        & 
 {4.0}\\
\hline
 {$\pi^\pm\pi^\mp$}                       & 
 {44}                                  & 
 {$10.0^{+6.8}_{-6.0}$}                   & 
 {2.2$\sigma$}                            & 
 {$0.7\pm 0.4$}                        & 
 {1.5}\\
 {$\pi^\pm\pi^0$}                         & 
 {37}                                       & 
 {$11.3^{+6.3}_{-5.2}$}                   & 
 {2.8$\sigma$}                            & 
 {$0.9^{+0.6}_{-0.5}$}                    & 
 {2.0}\\
 {$\pi^0\pi^0$ }  *                       & 
 {26}                                     & 
 {$1.2^{+1.7}_{-0.9}$}                    & 
 {}                                       & 
 {}                                       & 
 {0.9}\\
\hline
 {$K^\pm K^\mp$}                       & 
 {44}                                  & 
 {$0.0_{-0.0}^{+1.3}$}                 & 
 {0.0$\sigma$}                         & 
 {}                                    & 
 {0.4}\\
 {$K^\pm K^0$}                         & 
 {12}                                  & 
 {$0.6_{-0.6}^{+3.8}$}                 & 
 {0.2$\sigma$}                         & 
 {}                                    & 
 {2.1}\\
 {$K^0 \bar {K^0}$}                    & 
 {5}                                   & 
 {0}                                   & 
 {}                                    & 
 {}                                    & 
 {1.7}\\
\hline
 {$K^+\eta^{'}$}                     & 
 {5}                         & 
 {$12.0\pm 3.7$}                     & 
 {5.5$\sigma$}                               & 
 {$7.8^{+2.7}_{-2.2}\pm 1.0$}               & 
 {}\\
 {$\pi^+\eta^{'}$}                 & 
 {5}                               & 
 {$1.4^{+2.0}_{-1.0}$}             & 
 {2.0$\sigma$}                             & 
 {}                                & 
 {4.4}\\
 {$h^+\eta$}                       & 
 {9}                               & 
 {0}                               & 
 {}                                & 
 {}                                & 
 {0.8}\\
\hline
 {$h^\pm \pi^\mp$}                       & 
 {44}                                    & 
 {$31.7^{+8.4}_{-7.3}$}                  & 
 {7.8$\sigma$}                           & 
 {$2.2^{+0.6}_{-0.5}$}                   & 
 {}\\
 {$h^\pm \pi^0$}                         & 
 {37}                                    & 
 {$20.0^{+6.8}_{-5.9}$}                  & 
 {5.5$\sigma$}                           & 
 {$1.6^{+0.6+0.3}_{-0.5-0.2} \pm 0.1$}   & 
 {}\\
 {$h^\pm {K^0}$}                         & 
 {12}                                    & 
 {$9.8^{+4.5}_{-4.0}$}                   & 
 {4.4$\sigma$}                           & 
 {$2.4^{+1.1+0.2}_{-1.0-0.2} \pm 0.2$}   & 
 {}\\
\hline
\end{tabular}
\end{center}
\end{table}

In this paper we provide only a brief description of the analysis. 
Further details can be found in Refs.~\cite{rarebprd,fkwphd}.
Two kinematic variables, 
$\mathrm{M_B} = \sqrt{\mathrm{E^2_{beam}}-\mathrm{P^2_B}},
\mathrm{\ and\ }\Delta \mathrm{E}= \mathrm{E}_1 +
\mathrm{E}_2 -
\mathrm{E_{beam}}
$\  
are used to form a two dimensional
signal plus sideband region.
Using the beam energy in $M_B$\ 
improves the mass resolution
by an order of magnitude, resulting in $\sigma_{\mathrm{M_B}}\approx 2.6$MeV.

The main continuum background suppression is obtained by requiring
$|\cos\theta_{sph}|<0.8$. 
The angle $\theta_{sph}$\ here is the angle
between the candidate axis and the sphericity axis of the rest of the event.
Candidate $B$\ daughters from 
continuum background
tend to be the leading
particles in two back to back 5~GeV ``jets''. Background therefore
peaks towards $\cos\theta_{sph}=\pm 1$.
Signal is flat in $\cos\theta_{sph}$\ as
the two $B$'s are approximately at rest in the labframe, leading to
uncorrelated directions for the decay products.
This difference in ``event shape'' between $B\bar B$\ signal 
and continuum background is exploited further using a Fisher Discriminant
technique ($\cal F$) described in detail in Ref.~\cite{fkwphd}.
The yield is determined using a maximum likelihood fit for the fraction
of signal and background events out of the total number of events.
As input to the fit $M_B,\ \Delta E,\ {\cal F},\ \cos\theta_B$, and $dE/dx$\ 
information 
are used. The angle $\theta_B$\ is the $B$ decay angle with respect to the 
z-axis in the labframe. 
Decays of $\eta^{'}$\ are reconstructed in 
$\eta^{'}\to \eta\pi^+\pi^-
\to \gamma\gamma\pi^+\pi^-$. The search for $B^+\to h^+\eta$ includes
$\eta\to \gamma\gamma$ as well as $\eta\to \pi^+\pi^-\pi^0$.
The $\eta,\ \eta^{'}$\ mass is used as further 
input to the maximum likelihood fit where applicable. And $\cos\theta_B$\ 
is used as part of $\cal F$ rather than in the fit in those cases.

\begin{figure}
\centering
\leavevmode
\epsfysize=10cm
\epsfbox[47 115 477 538]{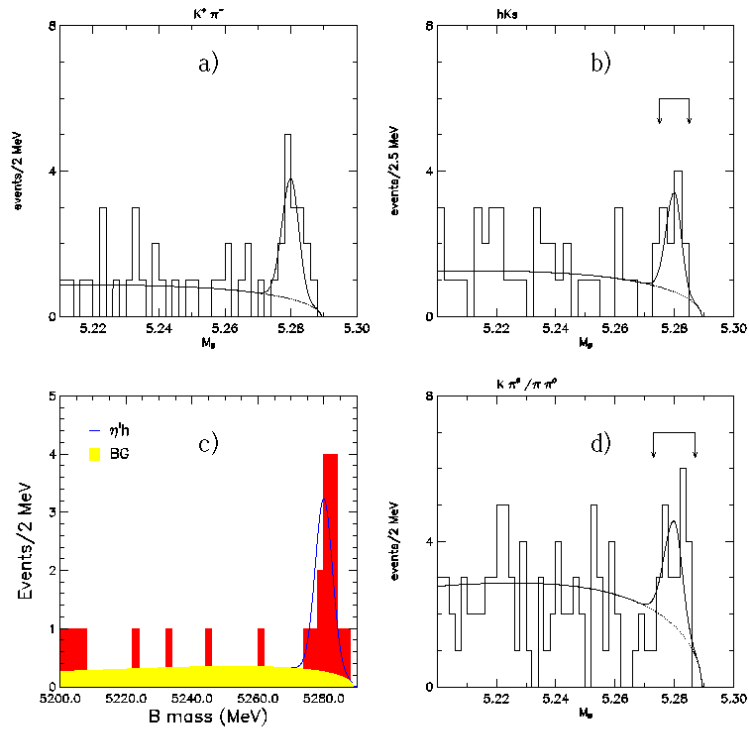}
\caption{
Mass distributions for $B$\ decays to 
a)$K^+\pi^-$, b)$K^0h^+$, c)$K^+\eta^{'}$\ and d)$h^+\pi^0$.
Projections from the fit are also shown. 
}
\label{fig:mass}
\end{figure}

Mass distributions for $B$\ decays to 
$K^+\pi^-,\ K^0h^+,\ K^+\eta^{'}$\ and $h^+\pi^0$\ are shown in 
Figure~\ref{fig:mass}. Additional cuts are applied
to suppress backgrounds in these plots. 
The curves are the PDF used in the fit normalized to the fit result
times the efficiency of the additional cuts applied.

\section{Discussion of Results}
\subsection{$\Delta S=1$\ Transitions}
We have measured the branching fractions for exclusive $B$\ decays to
the final states $K^+\pi^-,\ K^0\pi^+$\ and $K^+\eta^{'}$. All three of these
are $\Delta S = 1$\ transitions. 

It is very instructive to compare the square root of the three measured
branching fractions with each other, as well as the diagrams that
are expected to contribute.
For simplicity we ignore diagrams that are expected to be suppressed by
$O(\lambda^2)$. For completeness, we have also listed the
upper limits in $h^+\eta,\ K^0\pi^0$\ and the central value from the fit
in $K^+\pi^0$. 
\begin{equation}
\begin{array}{ccccc}
A_{K^0\pi^+} &=& (4.8^{+1.1}_{-1.0})\times 10^{-3}&\ \ \  & |P| \\
A_{K^+\pi^-} &=& (3.9^{+0.6}_{-0.5})\times 10^{-3}&\ \ \  & |T+P|\\
\sqrt{6}/3\times A_{K^+\eta^{'}}&=&(7.2^{+1.5}_{-1.2})\times 10^{-3}&\ \ \  &
{1\over 3} |T+3P+4P_1+(c_u + 3c_d) P_{EW}|\\
\sqrt{3}\times A_{K^+\eta} &<& 4.9\times 10^{-3}&\ \ \  & |T+P_1+c_uP_{EW}|\\
\sqrt{2}\times A_{K^0\pi^0} &<& 8.9\times 10^{-3}&\ \ \  &
|P -(c_u-c_d)P_{EW}| \\ 
\sqrt{2}\times A_{K^+\pi^0} &=& (3.7^{+1.1}_{-0.9})\times 10^{-3}&\ \ \  & 
|T+P+(c_u-c_d)P_{EW}|\\
\end{array}
\end{equation}

The amplitude $P_1$\ enters due to $\eta_1-\eta_8$\ mixing. 
It refers to the flavor singlet penguin diagram.
We have followed
Ref.~\cite{gronaueta} in our choice of mixing angle of 
$\phi = \sin^{-1}(1/3)\approx 20^\circ$. 
For this choice of $\phi$\ there is no flavor octet
contribution \PD\ in $B^+\to K^+\eta$. 
Varying this angle within its known range~\cite{etamixing} makes no 
difference
to the general arguments presented here.

The branching fraction in $K^0\pi^+$\ sets the scale by providing
a direct measurement of the
\PD\ amplitude. 
Measured branching fractions and upper limits for all other $\Delta S=1$\ 
transitions are
consistent with being dominated by the measured \PD\ amplitude.
In particular, we see no need to invoke
a $c\bar c$ or glueball component, nor anomalous coupling of two gluons
to $\eta^{'}$ in order to explain
the relative size of these branching fractions~\cite{wackoetaprime}.

Theoretical predictions of absolute branching fractions 
have large uncertainties due to factorization hypothesis and
poorly known form factors.
Keeping that in mind, theoretical predictions~
\cite{pengprediction,flpeng,charmpeng}
\ of
${\cal B}(B^+\to\pi^+K^0)\approx (1-2)\times 10^{-5}$\ 
agree surprisingly well with our experimental result.

Let us look at Eq.~(2) in some more detail. 
To $O(\lambda^4)$, the only non-trivial weak phases in the 
CKM-matrix are those of $V_{ub}$\ and $V_{td}$. 
The relative weak phase between
\TD\ and \PD\ for $\Delta S=1$\ transitions is thus the phase of
$V_{ub}$. 
The ratio of ${\cal B}(B^0\to K^+\pi^-)/{\cal B}(B^+\to K^0\pi^+)$\ 
may therefore provide constraints on the poorly
known phase of $V_{ub}$\ as was pointed out in Ref.~\cite{flpeng}.

The ratio of flavor singlet to flavor octet gluonic penguin diagrams 
$|P_1/P|$\ is rather difficult to estimate theoretically.
Neglecting \TD\ and \EWP , we find that our current upper limit
on ${\cal B}(B^+\to K^+\eta )$\ is consistent with the naive expectation
of $|P_1|< |P|$. 
This may provide a more stringent limit on $|P_1|$\ as we increase
our data set.
Similarly, a significant discrepancy between 
the ratio of branching fractions for 
$K^0\pi^+/K^+\pi^-$\ and $K^0\pi^0/K^+\pi^0$\ may in the future provide 
a lower limit on $|$\EWP $|$. 
Construction of an amplitude quadrangle for these
modes may in certain cases even provide information on the relative phases
of these amplitudes.

\subsection{$\Delta S=0$\ Transitions}
While we do see some excess of events above background in $\pi^+\pi^-$
and $\pi^+\pi^0$, the respective statistical significance of 
$2.2\sigma$\ and $2.8\sigma$\ is quite marginal.
Both of these decay modes are expected to be dominated by simple external
W-emission (\TD ) diagrams. Factorization may therefore be less questionable 
here than in the $\Delta S = 1$\ transitions discussed above.

Using the CLEO measurement 
${\cal B}(B^0\to\pi^- l^+\nu)=(2.0\pm 0.5\pm 0.3)\times 10^{-4}$
~\cite{pilnu}\  
we can use the factorization hypothesis, ISGW II, and the QCD factor
$a_1 = 1.01\pm 0.02$~\cite{a1}\ to predict the branching fractions
${\cal B}(B^0\to\pi^+\pi^-)= (1.3\pm 0.4)\times 10^{-5}$\ and
${\cal B}(B^+\to\pi^+\pi^0)= (0.7\pm 0.2)\times 10^{-5}$\ 
respectively~\cite{lkg}.
Uncertainties in the formfactor and factorization hypothesis are not
reflected in the errors quoted here. Furthermore,
contributions from anything other than the \TD diagram are neglected in this
kind of comparison. 
Keeping this in mind, we conclude that
the central value from the fit to the 
experimental data as shown in Table~\ref{tab:results} compares well with
these predictions. 

We do not see any evidence for $B^0\to\pi^0\pi^0$\ or $K^0\bar K^0$.
The dominant contributions to these decays are due to
($C - P$), and \PD\ respectively.
The penguin diagrams in both cases are $b\to d$\ penguins. 
Theoretical predictions for these modes range from less than $10^{-7}$\ to
few times $10^{-6}$\cite{pengprediction,charmpeng}.

Finally, we see no evidence for
$B^0\to K^+K^-$. 
This is not surprising as this decay can only proceed via \EX\ or
\PA\ diagrams. Theoretical predictions for this process are at the
level of at most a few times $10^{-8}$\cite{exchange}. 

We can therefore conclude that an overall consistent picture of
charmless hadronic $B$\ decays to two pseudoscalar mesons is starting
to emerge. 
CLEO has measured the dominant $\Delta S =1$\ transitions at levels
consistent with theoretical predictions. No signals are found in any
of the decay modes that are expected to be suppressed.

\section{Acknowledgments}
Many thanks to all the other people involved in the search for $B$\ decays to
two charmless pseudoscalar mesons at CLEO: Peter Gaidarev, Jim Alexander, 
Peter Kim, Andrei Gritsan, Jean Roy, and Jim Smith.
Further thanks go to Lawrence Gibbons for information 
concerning the factorization test, 
and to Robert Fleischer, Karl Berkelman and Jim Smith for many useful 
discussions.


\end{document}